\documentclass[aps,prb,preprint,showpacs]{revtex4}
\usepackage[T1]{fontenc}
\usepackage[latin1]{inputenc}
\usepackage{graphics}

\makeatletter

\providecommand{\LyX}{L\kern-.1667em\lower.25em\hbox{Y}\kern-.125emX\@}
\let\SF@@footnote\footnote
\def\footnote{\ifx\protect\@typeset@protect
    \expandafter\SF@@footnote
  \else
    \expandafter\SF@gobble@opt
  \fi
}
\expandafter\def\csname SF@gobble@opt \endcsname{\@ifnextchar[
  \SF@gobble@twobracket
  \@gobble
}
\edef\SF@gobble@opt{\noexpand\protect
  \expandafter\noexpand\csname SF@gobble@opt \endcsname}
\def\SF@gobble@twobracket[#1]#2{}

\usepackage[T1]{fontenc}
\usepackage[latin1]{inputenc}
\usepackage{graphics}

\makeatletter

\usepackage[T1]{fontenc}
\usepackage[latin1]{inputenc}
\usepackage{graphics}

\makeatletter

\usepackage[T1]{fontenc}
\usepackage[latin1]{inputenc}
\usepackage{graphics}

\makeatletter

\usepackage[T1]{fontenc}
\usepackage[latin1]{inputenc}
\usepackage{graphics}
\usepackage{graphicx,psfrag,color,pstcol,pst-grad}
\makeatletter

\usepackage[T1]{fontenc}
\usepackage[latin1]{inputenc}
\usepackage{graphics}
\usepackage[T1]{fontenc}
\usepackage[latin1]{inputenc}
\usepackage{graphics}
\usepackage[T1]{fontenc}
\usepackage[latin1]{inputenc}
\usepackage{graphics}
\usepackage[T1]{fontenc}
\usepackage[latin1]{inputenc}
\usepackage{graphics}
\usepackage{amsmath}

\makeatletter
\providecommand{\LyX}{L\kern-.1667em\lower.25em\hbox{Y}\kern-.125emX\@}
\let\SF@@footnote\footnote
\def\footnote{\ifx\protect\@typeset@protect
    \expandafter\SF@@footnote
  \else
    \expandafter\SF@gobble@opt
  \fi
}
\expandafter
\def\csname SF@gobble@opt \endcsname{\@ifnextchar[  \SF@gobble@twobracket
  \@gobble
}
\edef\SF@gobble@opt{\noexpand\protect
  \expandafter\noexpand\csname SF@gobble@opt \endcsname}
\def\SF@gobble@twobracket[#1]#2{}
\makeatletter
\providecommand{\LyX}{L\kern-.1667em\lower.25em\hbox{Y}\kern-.125emX\@}
\let\SF@@footnote\footnote
\def\footnote{\ifx\protect\@typeset@protect
    \expandafter\SF@@footnote
  \else
    \expandafter\SF@gobble@opt
  \fi
}
\expandafter
\def\csname SF@gobble@opt \endcsname{\@ifnextchar[  \SF@gobble@twobracket
  \@gobble
}
\edef\SF@gobble@opt{\noexpand\protect
  \expandafter\noexpand\csname SF@gobble@opt \endcsname}
\def\SF@gobble@twobracket[#1]#2{}
\makeatletter
\makeatletter
\providecommand{\LyX}{L\kern-.1667em\lower.25em\hbox{Y}\kern-.125emX\@}
\let\SF@@footnote\footnote
\def\footnote{\ifx\protect\@typeset@protect
    \expandafter\SF@@footnote
  \else
    \expandafter\SF@gobble@opt
  \fi
}
\expandafter
\def\csname SF@gobble@opt \endcsname{\@ifnextchar[  \SF@gobble@twobracket
  \@gobble
}
\edef\SF@gobble@opt{\noexpand\protect
  \expandafter\noexpand\csname SF@gobble@opt \endcsname}
\def\SF@gobble@twobracket[#1]#2{}
\input{tcilatex}
\makeatother

\makeatother

\makeatother

\begin{document}

\title{Magnetovibrational coupling in small cantilevers}

\author{Alexey A. Kovalev}

\affiliation{Department of NanoScience, Delft University of Technology, Lorentzweg
1, 2628 CJ Delft, The Netherlands}

\author{Gerrit E.W. Bauer}

\affiliation{Department of NanoScience, Delft University of Technology, Lorentzweg
1, 2628 CJ Delft, The Netherlands}

\author{Arne Brataas}

\affiliation{Department of Physics, Norwegian University of Science and Technology,
N-7491 Trondheim, Norway}

\begin{abstract}
A nano-magnetomechanical system consisting of a cantilever and a thin magnetic
film is predicted to display magnetovibrational modes, which should enable applications
for sensors and actuators. The {}``polaritonic{}'' modes can be detected by
line splittings in ferromagnetic resonance spectra. 
\end{abstract}


\date{\today{}}

\maketitle
Two unconventional roads into the realm of nano-scale devices may lead to useful
results in the not too distant future, \textit{viz}. magnetoelectronics\cite{Wolf:sc01}
and nano-electromechanical systems (NEMS).\cite{Roukes:pw01} In magnetoelectronics,
the spin degree of freedom of the electron is employed for new or extended functionalities,
which can be used \textit{e.g.} in random-access memories (MRAM),\cite{Parkin:amn02}
whereas advanced sensing and actuating have been realized by mechanical systems,
though yet limited to the micrometer scale (MEMS).\cite{Sidles:rmp95} With
the notable exception of mechanically detected ferromagnetic resonance (FMR),\cite{Jander:apl01}
these two fields have surprisingly little overlap. Here we demonstrate that
a combination of a conventional nano-mechanical structure, namely an elastic
bar clamped on one side and freely vibrating on the other ({}``cantilever{}''),
displays new physics when combined with a film of ferromagnetic material deposited
on top of it. We find theoretically that the lowest vibrational modes of the
cantilever are mixed with the coherent motion of the magnetic order parameter
by the magnetic anisotropy. We can translate information carried by rf (microwave) fields
at certain frequencies effectively into mechanical motion and \textit{vice versa.}
The present case study leads us to believe that nanoscale magneto-mechanical
systems deserve more attention.

Some time ago Benda \cite{Benda:ieeem69} predicted that FMR implies a mechanical
torque. Observation of its effect on the mechanical motion requires small samples,
however. Experimental proof of mechanical detection of FMR (as well as electron
paramagnetic resonance) has therefore been obtained only recently.\cite{Jander:apl01}
In these experiments the torque is detected by a static deflection of a cantilever
in the presence of DC and AC magnetic fields. Here we study a similar system
(sketched in Fig. \ref{fig1}) but concentrate on the high frequency regime,
in which the magnetic and mechanical motion turn out to be coupled. Traditionally
time scales of FMR (usually at GHz frequencies) and mechanical oscillations
(in MEMS typically KHz) are different and this mode mixing cannot be observed.
More recently, NEMS oscillate at MHz frequencies or even GHz,\cite{Huang:nat03}
which is accessible by FMR. The predicted effects can be measured for example
by calorimetric techniques \cite{Moreland:rsi00} or, indeed, mechanically.\cite{Jander:apl01}

We consider a small dielectric cantilever with a single-domain ferromagnetic
layer deposited on its far end (see Fig. \ref{fig1}). A constant external field
\( \mathbf{H}_{0} \) is applied along the \( y \) axis as well as an oscillating
field \( H_{\text {x}} \) along the \( x \) axis. The effective field \( \mathbf{H}_{\text {eff}} \)
felt by the magnetization consists of \( \mathbf{H}_{0} \) as well as the crystal
anisotropy and the demagnetizing fields. The strains are localized in the mechanical
link of the cantilever between the ferromagnetic film at one end and the other
end that is fixed. The lattice of the ferromagnet then oscillates without internal
mechanical strains, but crystalline and form anisotropies couple the magnetic
order parameter to the torsional mode of the cantilever (the longitudinal mode
does not cause any internal magnetization dynamics).

The magnetization \( \mathbf{M} \) of the ferromagnet precesses around an effective
magnetic field \( \mathbf{H}_{\text {eff}} \) according to the Landau-Lifshitz-Gilbert
equation:\cite{Gilbert:pr55}\begin{equation}
\label{LL+current}
\frac{d\mathbf{M}}{dt}=-\gamma \mathbf{M}\times \mathbf{H}_{\text {eff}}+\alpha \mathbf{M}\times \frac{d\mathbf{M}}{dt},
\end{equation}
 where \( \gamma  \) denotes the gyromagnetic ratio. The phenomenological Gilbert
constant is typically \( \alpha \leq 0.01 \). It will be disregarded in the
following for simplicity, which is valid when the mechanical damping dominates.
To first order, the deviations from the equilibrium magnetization in the \( y \)-direction
lie in the \( x-z \) plane: 
$\mathbf{M}=m_{\text {x}}\mathbf{x}+M_{\text {s}}\mathbf{y}+m_{\text {z}}\mathbf{z}$,
 where \( M_{\text {s}} \) is the saturation magnetic moment. Without loss
of generality we may take the crystalline anisotropy field contribution \( H_{\text {A}} \)
to be uniaxial and directed along the \( y \)-axis of the system at rest. \( H_{\text {A}} \)
rigidly follows the vibration of the crystal lattice described by the torsion
angle \( \varphi  \) (see Fig. 1). For small \( \varphi  \), the effective
field oscillates in the \( y-z \) plane:
$\mathbf{H}_{\text {eff}}=(H_{\text {A}}+H_{0})\mathbf{y}+\left( H_{\text {A}}\varphi +\nu M_{\text {s}}\varphi -\nu m_{\text {z}}\right) \mathbf{z}$,
 where \( \nu  \) describes the demagnetizing dipolar field (\( \nu \simeq 4\pi  \)
for our geometry). Note that the coupling does not rely on a strong magnetocrystalline
field \( H_{\text {A}}, \) since the surface forces of demagnetizing currents
also provide a restoring torque. For a thin ferromagnetic layer the latter dominates
and the magnetization does not precess, but oscillates like a pendulum in the
\( x-y \) plane due to the oscillating field in the \( z \)-direction \( \mathbf{H}_{\text {eff}}=\nu M_{\text {s}}\varphi \mathbf{z} \).

The torsional motion of the part of the cantilever that is not covered by the
ferromagnet can be found applying the variational principle to the total elastic
energy:\cite{Landau:59}\begin{equation*}
E_{el}=\frac{1}{2}\int_{0}^{L}C\tau ^{2}dx ,
\end{equation*}where \( \tau =\partial \varphi /dx\, \,  \)and \( C \) is
an elastic constant defined by the shape and material of the cantilever (\( C=\frac{1}{3}\mu da^{3} \)
for a plate with thickness \( a \) much smaller than width \( d \), \( a\ll d \),
\( \mu  \) is the Lam\( \acute{\mathrm{e}} \) constant). \( T=C\tau \left( x\right)  \)
is the torque flowing through the cantilever at point \( x \). The integration
is taken from the clamping point \( x=0 \) until the cantilever endpoint \( x=L \).
The equation of motion reads\begin{equation}
C\frac{\partial ^{2}\varphi }{\partial x^{2}}=\rho I\frac{\partial ^{2}\varphi }{\partial t^{2}}+2\beta \rho I\frac{\partial \varphi }{\partial t},
\end{equation}
 where \( I=\int (z^{2}+y^{2})dzdy\simeq ad^{3}/12 \) is the moment of inertia
of the cross-section about its center of mass, \( \rho  \) the mass density,
and \( \beta  \) is a phenomenological damping constant related to the quality
factor \( Q \) at the resonance frequency \( \omega _{\text {e}} \) as \( Q=\omega _{\text {e}}/(2\beta ) \).
The oscillating solution has the form
$\varphi =\sin (kx)(A_{1}\sin (\omega t)+A_{2}\cos (\omega t))$,
 where \( k=\omega /c \) is the wave number, \( c=c_{\text {t}}2h/d \) and
\( c_{\text {t}}=\sqrt{\mu /\rho } \) is the transverse velocity of sound.
The free constants \( A_{1} \) and \( A_{2} \) depend on the initial conditions.
The clamping boundary condition \( \varphi |_{x=0}=0 \) at the clamping point
is already fulfilled, and the boundary condition at the end \( x=L \) is discussed
in the following.

The conservation law for the mechanical angular momentum \( \mathbf{V}^{\text {el}} \)
for \( x\in \left\{ 0,L\right\}  \) (without magnetic overlayer) \( d\mathbf{V}^{\text {el}}\left( x\right) /dt=\mathbf{T}\left( x\right) ,\, \,  \)where
the torque \( \mathbf{T} \) has been defined above, is modified by the coupling
to the magnet in a region \( x\in \left\{ L,L+\Delta L\right\}  \) (\( \Delta L \)
is the length of the cantilever covered by the magnetic layer) as:\begin{equation}
\label{motion}
\frac{d}{dt}\left( \mathbf{V}^{\text {el}}\left( x\right) +(-\frac{\hbar }{2\mu _{\text {B}}})\mathbf{M}\left( x\right) \right) =\mathbf{T}\left( x\right) +\mathbf{T}_{\text {field}}.
\end{equation}
 where \( \mathbf{T}_{\text {field}}=\frac{\hbar }{2\mu _{\text {B}}}\gamma \mathbf{M}\times \mathbf{H}_{0} \)
(\( \mu _{\text {B}} \) is Bohr magneton). For small amplitudes, \( V_{\text {x}}^{\text {el}}=N\frac{d}{dt}\varphi  \)
and \( T_{\text {x}}=-K\varphi  \), where \( N \) and \( K \) are geometry
and material specific constants. Assuming \( \Delta L\ll L, \) internal strains
in the magnetic section \( x\in \left\{ L,L+\Delta L\right\}  \) are disregarded.
The magnetovibrational coupling may then be treated as a boundary condition
to the mechanical problem, which is expressed as the torque \( C\tau |_{x=L} \)
exerted by the magnetization on the cantilever: \begin{equation}
\label{bound}
C\tau |_{x=L}=-\frac{\hbar }{2\mu _{\text {B}}}\gamma \mathbf{M}\times \left( \mathbf{H}_{\text {eff}}-\mathbf{H}_{0}\right) |_{x}-N\frac{d^{2}}{dt^{2}}\varphi ,
\end{equation}
 By equating this with the substitution of $\varphi$ in $\tau$,
$ C\tau |_{x=L}=Ck\cos (kL)(A_{1}\sin (\omega t)+A_{2}\cos (\omega t))=Ck\varphi \cot (kL) $,
we obtain in frequency space\begin{eqnarray}
Ck\varphi \cot (kL) & = & \frac{\hbar }{2\mu _{\text {B}}}(-i\omega m_{\text {x}}+\gamma H_{0}m_{\text {z}})-\omega ^{2}N\varphi \label{Fourier} \\
 & \approx  & -C\varphi \frac{L}{2c^{2}}(\omega ^{2}+2i\beta \omega -\omega _{\text {e}}^{2}),\label{Fourierexp} 
\end{eqnarray}
 where in the second line the \( \cot  \) has been expanded close to the resonance
frequency \( \omega _{\text {e}}=c\pi(1/2+s) /L  \) ( \(s\) is integer, here we concentrate mainly on coupling to the first harmonic and \(s=0\)). For negligible
damping of the magnetization dynamics, the external rf field \( H_{\text {x}} \)
does not create any torque along the \( x \) axis. \( N \) scales like \( \Delta L/L \)
and we therefore disregard the term proportional to \( N \) (as long as \( N\ll CLd^{2}/2c^{2} \)
it only slightly shifts the resonant frequency \( \omega _{\text {e}} \)).

The magnetic susceptibility \( \chi _{\omega }=\left( m_{\text {x}}/H_{\text {x}}\right) _{\omega } \)
describes the linear response of the magnetization \( m_{\text {x}} \) to a
(weak) rf magnetic field \( H_{\text {x}} \) at frequency \( \omega  \) and
can be found after writing Landau-Lifshitz equation in frequency space with
use of Eq. (\ref{Fourier}): \begin{equation}
\label{suscept}
\frac{\chi _{\omega }}{\gamma ^{2}M_{\text {s}}V}=\left[ \frac{\omega ^{2}-\omega _{\text {m}}^{2}}{(H_{\text {A}}+H_{0}+\nu M_{\text {s}})}+\frac{\omega ^{2}GL\tan (kL)}{2kc^{2}(H_{\text {A}}+\nu M_{\text {s}}+H_{0}(1-GL\tan (kL)/2kc^{2}))}\right] ^{-1}
\end{equation}
 where \( G=\hbar M_{\text {s}}V\gamma (H_{\text {A}}+\nu M_{\text {s}})c^{2}/\left( \mu _{\text {B}}CL\right)  \)
is the magnetovibrational coupling constant (\( V \) is the volume of the magnet)
and the unperturbed resonance frequency \( \omega _{\text {m}}^{2}=\gamma (H_{\text {A}}+\nu M_{\text {s}})\gamma H_{\text {A}} \).
The imaginary part of \( \chi _{\omega } \) is proportional to the FMR\ absorption
signal. In the absence of the external field \( H_{0} \) the resonance frequencies
in the vicinity of the resonance frequency \( \omega _{\text {e}} \) can be
found after expanding \( \cot (kL) \) as in Eq. (\ref{Fourierexp}):\begin{equation}
\label{resonance}
\omega _{1(2)}=\sqrt{\frac{1}{2}}\left[ \omega _{\text {e}}^{2}+\omega _{\text {m}}^{2}+G\pm \left( ((\omega _{\text {e}}+\omega _{\text {m}})^{2}+G)((\omega _{\text {e}}-\omega _{\text {m}})^{2}+G)\right) ^{1/2}\right] ^{1/2}.
\end{equation}
When the external field \( H_{0} \)
is smaller than the demagnetizing field, Eq. (\ref{resonance}) holds with \( \omega _{\text {m}}^{2}=\gamma ^{2}(H_{\text {A}}+H_{0}+\nu M_{\text {s}})(H_{\text {A}}+H_{0}) \).
This fact can be used to tune the FMR frequency in order to match the elastic
frequency. In this limit and \( H_{0}A/(M_{\text {s}}\beta \omega _{\text {e}})\ll 1, \)
Eq. (\ref{suscept}) simplifies to\begin{equation}
\label{suscept1}
\chi _{\omega }\approx \frac{\gamma ^{2}\nu M_{\text {s}}V(H_{\text {A}}+\nu M_{\text {s}})}{\omega ^{2}-\omega _{\text {m}}^{2}+\omega ^{2}GL\tan (kL)/\left( 2kc^{2}\right) }.
\end{equation}
 Its imaginary part corresponding to the rf absorption is plotted in Fig. \ref{fig2}
for \( G>0 \) (which is the case in our setup). We observe a typical anticrossing
behavior between an optically active and non-active mode, with level repulsion
and transfer of oscillator strength. The intrinsic damping of the mechanical
system (for MEMS \( Q \) factors can reach \( 10^{4}, \) which quickly deteriorate
with decreasing size, however) imposes an extra damping on the magnetization
dynamics, which close to the mode crossing may dominate the intrinsic damping
due to a small Gilbert constant \( \alpha  \) disregarded here.

The novel magnetovibrational coupling in our cantilever is observable by FMR,
but the signal of nanoscale magnets is small. It might therefore be preferable
to detect the resonance by the static deflection of the same cantilever due
to an additional constant magnetic field \( \mathbf{H}_{T} \) along the \( z \)-axis,
as described by Lohndorf \textit{et al}.\cite{Lohndorf:apl00}. In our approximation
the field \( \mathbf{H}_{\text {T}} \) creates a torque \( \gamma M_{\text {y}}H_{\text {T}}\vec{x} \),
whose modulation at the FMR conditions should be detectable.

Since the vibrational frequencies of state-of-the-art artificial structures
are relatively low, the use of soft ferromagnets (such as permalloy), is advantageous.
The magnetic mode frequencies are then determined by shape anisotropies. The
FMR frequency and the mechanical resonance frequency should not differ by much
more than \( \Delta \omega \sim \sqrt{G} \) for a pronounced effect. A Si cantilever
with \( a\times d\times L=\left( 1\times 5\times 50\right) \mu m \) \( \left( C=10^{-13}Nm^{2}\right)  \)
has a torsional resonance frequency of the order of \( \omega _{\text {e}}=10 \)
MHz. Taking our ferromagnetic layer of dimensions \( a1\times d\times \Delta L=50nm\times 5\mu m\times 5\mu m \)
(thickness, width and length), then \( \sqrt{G}\sim 100\, \, KHz \), meaning
that we should tune the magnetic resonance to \( \omega _{\text {e}}\pm 100\, \, KHz \)
to observe the \textquotedblleft polariton\textquotedblright. The necessary
rf field \( H_{\text {x}} \) depends on the elastic and magnetization viscous
dampings. At low frequencies additional sources of damping complicate
measurements\cite{Cochran:prb89} and the coherent motion of the magnetization
can be hindered by domain formation. Coupling to higher resonance modes\cite{Rabe:rsi96}
or structuring of the ferromagnet may help to carry out measurements. Downscaling
the device size is most favorable, since the mechanical resonance frequency
increases inversely proportional to the scale coefficient, whereas the coupling
constant scales as the second power of the inverse coefficient.

An actual observation of the predicted splittings would give information about
\textit{e.g.} the magnetic moment of the film and the broadening would yield
the quality factor of the elastic motion. From a technological point of view
the tunable damping due to the magnetovibrational coupling might be interesting
for optimizing switching speeds. A ferromagnet effectively absorbs microwaves
and turns them into a precessing magnetization, which via the magnetovibrational
coupling can be transformed into a coherent mechanical motion. On the other
hand, the ferromagnet may interact with the mechanical motion, to cause a magnetization
precession, which in turn emits polarized microwaves. The emission in the coupled
regime is more energy efficient in comparison with a fixed magnetic dipole emission
in the case of small Gilbert constant but relatively low mechanical quality
factor. This might be interesting for \textit{e.g}. on-chip communication applications. 

The present device can be interpreted as a MASER source in which the microwave
cavity is replaced by the mechanical resonator. After submission of the present
manuscript (cond-mat/0303114), Bargatin and Roukes (cond-mat/0304605) worked
out very similar ideas for nuclear spin systems. The magnetization reversal
of the ferromagnetic tip for parameters mentioned above can be described by
semiclassical Bloch-Maxwell laser equations \cite{Svelto:98} with coupling
constant \( g\sim 10\: s^{-1} \) and relaxation rate \( \Gamma \sim \omega _{\text {m}}\alpha  \),
as will be explained in more detail elsewhere.

To summarize, we have calculated the magnetic susceptibility of a system with
magnetovibrational coupling by magnetocrystalline fields or via surface forces
of demagnetizing currents. A condition for effective energy transfer from an
external rf magnetic field into mechanical motion and \textit{vice versa} has
been established. FMR spectra are predicted to split close to the resonance,
and to strongly depend on the mechanical damping. The predicted effects should
be observable with existing technology, but a further reduction of system size
would strongly enhance them.

We would like to acknowledge helpful discussions with Yaroslav Tserkovnyak and
interesting suggestions from the anonymous Referee. This work was supported
by the FOM and the NEDO project ``NAME''.

\newpage

\begin{figure}
\caption{A nano-magneto-mechanical cantilever supporting magneto-vibrational modes.
On a dielectric substrate (such as Si) a single-domain ferromagnetic film is
deposited at the free end.}

\label{fig1}
\end{figure}

\begin{figure}
\caption{a) Dependence of the resonance frequency of the coupled motion on the FMR frequency
\protect\( \omega _{\text {m}}\protect \)
of the uncoupled magnetization and b) the corresponding oscillator strength
of each resonance in arb. units plotted by full lines and its width plotted
by dashed lines (\protect\( \omega _{\text {e}}=10\protect \protect \)
MHz, \protect\( \Delta \protect \omega \sim 500\, \, KHz\protect \), \protect\( \alpha \protect =0\protect \)).}

\label{fig2}
\end{figure}

\newpage

\centerline{\includegraphics[width=7cm]{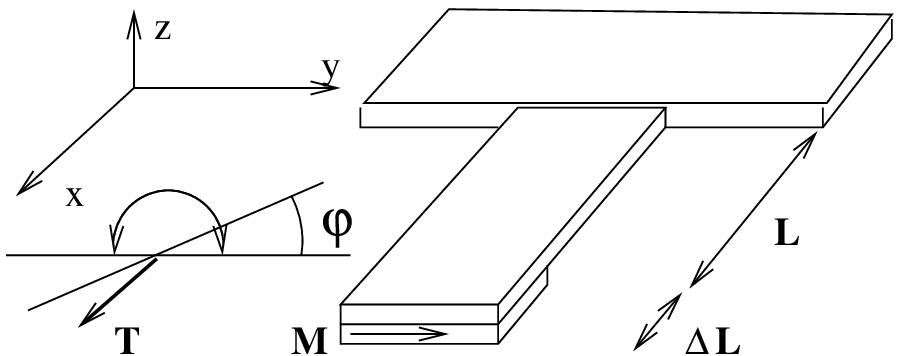}  }
\vspace{1cm}
\centerline{\Large{Fig. 1}}
\newpage
\centerline{ \includegraphics{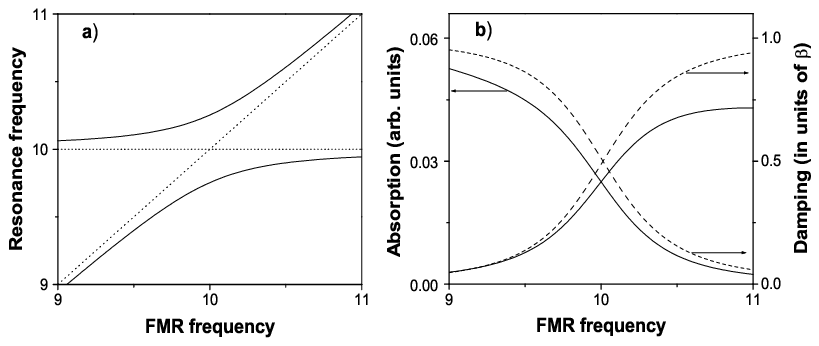} }
\vspace{1cm}
\centerline{\Large{Fig. 2}}
\end{document}